\documentclass[aps,prd,10pt,twocolumn,preprintnumbers,superscriptaddress,showpacs,floatfix,nofootinbib]{revtex4-2}
\usepackage{textcomp,gensymb}
\usepackage{hyperref}
\usepackage{graphicx}
\usepackage{amsfonts,amsmath,amssymb,bm,bbm}
\usepackage{color,soul}
\usepackage[dvipsnames]{xcolor}
\usepackage{slashed}
\usepackage{flushend}
\usepackage{physics}
\usepackage{graphicx}
\usepackage{dcolumn}
\usepackage{float}
\usepackage{comment}
\usepackage{tikz}
\usepackage{subcaption}
\usepackage[normalem]{ulem}
\usepackage[compat=1.1.0]{tikz-feynman}
\usepackage[utf8]{inputenc}

\usepackage{ragged2e}
\usepackage{subcaption}

\hypersetup{
    pdfnewwindow=true,      
    colorlinks=true,       
    linkcolor=blue,          
    citecolor=blue,        
    filecolor=blue,      
    urlcolor=blue        
}
\usepackage{graphicx}

\begin{document}
\preprint{FERMILAB-PUB-26-0090-PPD}
\preprint{MI-HET-879}

\title{Prying Open the Dark Sector Window with SBND Off-Target Mode}

\author{Bhaskar Dutta}\email{dutta@tamu.edu}\affiliation{Texas A$\&$M University, College Station, TX 77843, USA}
\author{Debopam Goswami}\email{debopam22@tamu.edu}\affiliation{Texas A$\&$M University, College Station, TX 77843, USA}
\author{Aparajitha Karthikeyan}\email{aparajitha$\_$96@tamu.edu}\affiliation{Texas A$\&$M University, College Station, TX 77843, USA}
\author{Vishvas Pandey}\email{vpandey@fnal.gov}\affiliation{Fermi National Accelerator Laboratory, Batavia, IL 60510, USA}
\author{Zahra Tabrizi}\email{z$\_$tabrizi@pitt.edu}
\affiliation{PITT PACC, Department of Physics and Astronomy, University of Pittsburgh, 3941 O'Hara St., Pittsburgh, PA 15260, USA}
\affiliation{Theoretical Physics Department, CERN, 1 Esplanade des Particules, CH-1211 Geneva 23, Switzerland}
\author{Adrian Thompson}\email{a.thompson@northwestern.edu}\affiliation{Northwestern University, Evanston, IL 60208, USA}
\author{Richard G.~Van de Water}\email{vdwater@lanl.gov}\affiliation{Los Alamos National Laboratory, Los Alamos, NM 87545, USA}

\begin{abstract}

Accelerator-based neutrino experiments with high-intensity proton beams and advanced detector technologies provide a powerful and complementary approach to probing physics beyond the Standard Model. The MiniBooNE experiment at Fermilab pioneered a dedicated Booster Neutrino Beam (BNB) off-target (beam-dump) run, setting leading constraints on sub-GeV dark matter. In this work, we explore the physics opportunities enabled by operating the Short-Baseline Near Detector (SBND) at Fermilab in a future BNB off-target configuration, as well as in a dedicated beam-dump configuration. By redirecting the proton beam away from the nominal beryllium target, or by employing a dedicated beam-dump, neutrino-induced backgrounds are substantially suppressed, thereby enhancing SBND’s sensitivity to many new physics scenarios. We demonstrate that such running modes significantly extend the reach for new physics. As representative examples, we present projected sensitivities to light dark matter, axion-like particles, heavy neutral leptons, and meson-portal scenarios.

\end{abstract}

\maketitle

%-----------------------------------------------------------------------%
\section{Introduction}
\label{sec:introduction}

Theoretical developments over the past decade have established dark sectors -- collections of particles that do not couple directly to the Standard Model (SM) through the strong, weak, or electromagnetic interactions -- as a compelling framework for addressing dark matter and other open questions in particle physics. Portal models, which connect SM particles to dark sector states via vector, scalar, or neutrino interactions, provide a systematic and experimentally testable pathway to probe these scenarios.

Accelerator-based neutrino beam facilities are particularly well suited for exploring such models due to their intense proton beams and large, high-resolution detectors~\cite{Acero:2022wqg,Karagiorgi:2022fgf}. At Fermilab, the Booster Neutrino Beamline (BNB) produces abundant secondary mesons from high-intensity proton collisions on target. These mesons can serve as sources of light mediators that subsequently decay into dark sector particles. Such particles may propagate to downstream detectors, including the Short-Baseline Near Detector (SBND), where they can interact or decay to produce observable signatures. SBND, located 110~m from the BNB target, is a 112-ton liquid argon time projection chamber (LArTPC) with excellent spatial and calorimetric resolution~\cite{MicroBooNE:2015bmn, SBND:2025lha}. 

A particularly promising strategy involves operating SBND in a beam-dump configuration, in which the proton beam is redirected away from the nominal beryllium target and magnetic horn system, referred to as the {\it off-target mode}. In this configuration, the production of neutrinos from charged meson decays is substantially reduced, while neutral meson production remains significant. As a result, neutrino-induced backgrounds can be suppressed by up to orders of magnitude, dramatically enhancing sensitivity to rare interactions involving light dark matter and other exotic particles. This strategy was pioneered by the MiniBooNE experiment, which conducted a dedicated BNB off-target (beam-dump) run in 2014 and set leading constraints on sub-GeV dark matter by suppressing neutrino-induced backgrounds~\cite{MiniBooNE:2017nqe,MiniBooNEDM:2018cxm}.

As the Fermilab accelerator complex prepares for upgrades under the PIP-II program, which will substantially increase the available beam power, the physics reach of SBND in beam-dump operation will correspondingly expand. In this work, we investigate the opportunities enabled by future off-target and dedicated beam-dump running of SBND following the planned accelerator shutdown and restart~\cite{SBND:2025lha}. We demonstrate that such a program can provide powerful and complementary sensitivity to a broad class of physics scenarios beyond the Standard Model. Importantly, beam-dump operation can in principle coexist with standard (anti)neutrino running through pulse-by-pulse selection of on-target or off-target beam delivery~\cite{Arrington:2022pon,Toups:2022knq,SBND:2025lha}. This flexibility would allow SBND to maximize its physics impact by simultaneously advancing both neutrino physics and searches for weakly coupled new particles. 

New physics signals with final states such as $e$, $\gamma$, $\gamma\gamma$, and $\pi^0$ benefit significantly from the reduced neutrino-induced backgrounds in beam-dump operation. The resulting cleaner environment enables substantially enhanced sensitivity to light dark matter (DM), axion-like particles (ALPs), heavy neutral leptons (HNLs), and related dark sector scenarios.

The remainder of this paper is organized as follows. In Sec.~\ref{sec:beamdump}, we describe the BNB operation scenario in on-target, off-target and dedicated beam-dump configurations. Then, we present the physics case for a variety of phenomenological dark sector models, motivated by dark matter, axion-like particles, heavy neutral leptons, and probes of short-baseline anomalies in Sec.~\ref{sec:dm}, Sec.~\ref{sec:alp}, Sec.~\ref{sec:hnl}, and Sec.~\ref{sec:meson}. We summarize our results and discuss their implications in Sec.~\ref{sec:summary}.

%-----------------------------------------------------------------------%
\section{Beam Operation Scenarios}
\label{sec:beamdump}

Proton beam-dump configurations provide intense sources of secondary mesons and constitute a powerful approach for probing weakly coupled new physics. In 2014, the MiniBooNE experiment performed a dedicated off-target (beam-dump) run in which neutrino-induced backgrounds were significantly suppressed, leading to competitive constraints on sub-GeV dark matter~\cite{MiniBooNE:2017nqe,MiniBooNEDM:2018cxm}. These results demonstrated the potential of beam-dump operation at neutrino facilities for dark sector searches. When a high-intensity proton beam is steered into a dense absorber rather than the nominal production target, the flux of neutrinos from charged meson decays is substantially reduced. In contrast, the production of short-lived neutral mesons, such as $\pi^0$ and $\eta$, remains sizable. These neutral mesons can decay into light mediators that subsequently produce dark sector particles capable of traversing the baseline and interacting in a downstream detector through rare but observable processes.

SBND provides a natural extension of this strategy. As a 112-ton liquid argon time projection chamber located 110~m from the BNB target, it combines proximity to the beam with excellent capabilities of the LArTPC technology. Following its nominal three-year neutrino-mode exposure of $1\times10^{21}$ POT, and after the planned accelerator restart, SBND could operate in an off-target or a dedicated beam-dump configuration~\cite{SBND:2025lha}. Moreover, beam-dump running can in principle be interleaved with standard (anti)neutrino operation through pulse-by-pulse selection of on-target or off-target beam delivery~\cite{Arrington:2022pon,Toups:2022knq,SBND:2025lha}. In this work, we consider three operational scenarios: 

\begin{itemize}
    \item \textbf{$\nu$-Mode}: In the nominal neutrino-mode (on-target mode) configuration, the proton beam impinges on the Be target. SBND is expected to accumulate $10^{21}$~POT over its three-year data-taking period as part of the SBN program.
    \item \textbf{Off-Target Mode}: In this configuration, the proton beam bypasses the Be target and magnetic horn system and is directed onto the existing 50-m Fe absorber. This setup suppresses the neutrino induced background by approximately a factor of 50~\cite{Toups:2022knq}, resulting in a substantially cleaner background environment while utilizing the existing accelerator and detector infrastructure. We consider three representative exposures: (i) $2\times10^{20}$~POT, corresponding to a potential run prior to the long shutdown; (ii) $4\times10^{20}$~POT, corresponding to a two-year run after the shutdown; and (iii) $6\times10^{20}$~POT, representing the combined exposure of the previous two scenarios.
    \item \textbf{Dedicated Beam-Dump Mode:} A more ambitious option involves constructing a dedicated beam-dump (BD) composed of dense material such as iron or tungsten. Placing this new absorber approximately 110~m upstream of SBND would suppress neutrino backgrounds by up to three orders of magnitude relative to standard on-target operation, significantly enhancing sensitivity to weakly coupled new physics. Simulations indicate that this configuration reduces decay-in-flight neutrino production while increasing the yield of neutral mesons that can decay into dark sector particles~\cite{Toups:2022knq}. In particular, the $\pi^0$ production rate per POT exceeds that of the standard target configuration (see Appendix~\ref{sec:fluxes}). We consider a three-year dedicated run after the long shutdown, corresponding to $1.8\times10^{21}$~POT. %A comparison plot between the target and dump produced fluxes of photons and mesons ($\pi^\pm$, $\pi^0$, $\eta$ and $K^+$) are shown in Fig.~\ref{fig:fluxplot}.
\end{itemize}

\begin{table*}[!htbp]
    \centering
    \renewcommand{\arraystretch}{1.2}
     \begin{tabular}{ | c | c | c | c | c | c | c | } 
%    \begin{tabular}{ | c | c | c | c | c |>{\centering\arraybackslash}p{2.5cm}| >{\centering\arraybackslash}p{3cm}| } 
        \hline
         Configuration & POT & Target & $\pi^0$  & $\eta$ & Background Counts &  Target distance \\
                       &  &  Material & per POT & per POT & (e-,~p,~$\pi^0$) &  from detector (m) \\
        \hline
        $\nu$-Mode (3 Years) & $1\times10^{21}$ & Be & 0.8825 & 0.0521 & (500, 2077379, 543095) & 110 \\ 
        \hline
        Off-Target Mode (1 Year) & $2\times10^{20}$ & Fe & 1.9986 & 0.0699 & (4, 27929, 7302) & 60\\ 
        \hline
        Off-Target Mode (2 Years) & $4\times10^{20}$ & Fe & 1.9986 & 0.0699 & (8, 55858, 14604) & 60 \\
        \hline
        Off-Target Mode (3 Years) & $6\times10^{20}$ & Fe & 1.9986 & 0.0699 & (12, 83787, 21906) & 60 \\
        \hline
        Dedicated Beam-Dump Mode (3 Years) & $1.8\times10^{21}$ & Fe & 1.9986 & 0.0699 & (0.9, 3739, 978) & 110 \\
        \hline
    \end{tabular}
    \caption{Mode specifications for SBND experiment used in this analysis, including POT values, $\pi^0$ and $\eta$ production rates from \texttt{GEANT4}~\cite{GEANT4:2002zbu}, background counts, and distance from the detector. Background counts in the top row for the $\nu$-mode are taken from Ref.~\cite{MicroBooNE:2015bmn}. Counts for the other modes are derived in this work.}  
    \label{tab:ExperimentalDetails_1}
\end{table*}

\begin{figure*}
    \centering
        \includegraphics[width=0.65\linewidth]{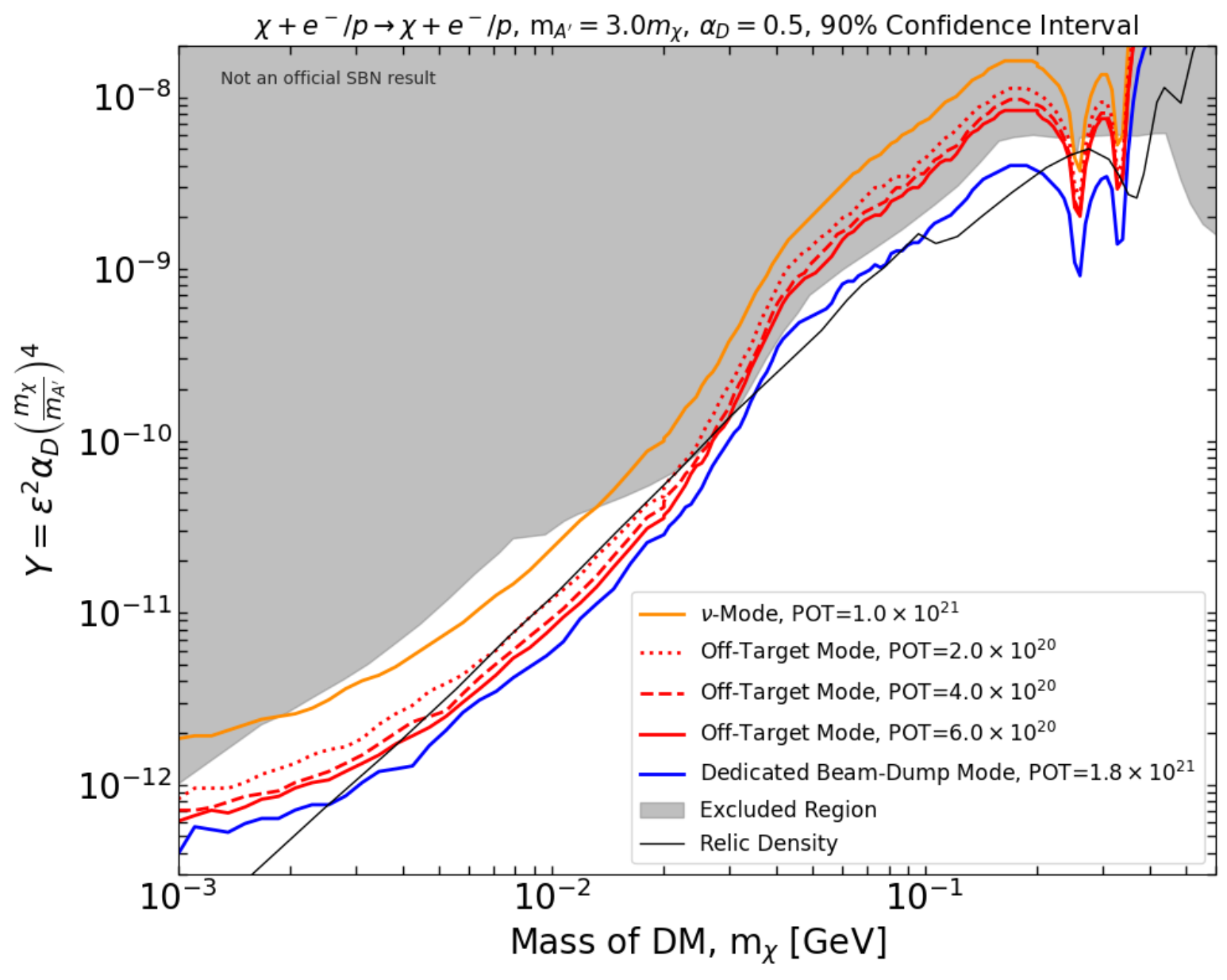}
%    \captionsetup{justification=Justified}
    \caption{Sensitivity to scalar DM at the 90\% confidence level for a vector mediator model with m$_{A^{'}}$$=3m$$_\chi$ and $\alpha_D=0.5$. The lines represent the projected sensitivities for different beam configurations at SBND. The DM is produced via meson decays and proton bremsstrahlung, with detection through DM-electron elastic scattering, DM-proton elastic scattering, and inelastic nucleon interactions. Existing constraints from BaBar, LSND, E137, COHERENT CsI, MiniBooNE, NA64, and direct detection experiments are included in the shaded gray region called ``Excluded Region", and the black line indicates the parameters satisfying the thermal relic abundance.}
    \label{fig:beam_dump}
\end{figure*}

We simulate the SM neutral meson ($\pi^0$ and $\eta$) fluxes~\cite{Capozzi:2021nmp,Capozzi:2024pmh, Dutta:2026zpe} using the \texttt{GEANT4}~\cite{GEANT4:2002zbu} simulation package. The resulting $\pi^0$ and $\eta$ yields per POT for the different configurations, along with the corresponding nominal background levels, are summarized in Table.~\ref{tab:ExperimentalDetails} and in Appendix.~\ref{sec:fluxes}. These configurations will also benefit from SBND's excellent timing resolution~\cite{SBND:2024vgn, Dutta:2025npn}, allowing potential identification of delayed, slow-moving dark sector signals relative to prompt neutrino interactions. Furthermore, anticipated increases in beam intensity under the PIP-II upgrade could allow high-statistics data collection over comparatively short timescales.

These dedicated operations scenarios therefore provide a favorable environment for exploring a broad class of weakly coupled new physics scenarios, including  dark matter, axion-like particles, heavy neutral leptons, and meson-portal scenarios. This strategy complements other experimental probes, including underground direct-detection experiments, astrophysical searches using cosmic-ray and gamma-ray observations, neutrino telescopes, and high-energy collider experiments. In the following sections, we present representative scenarios to illustrate the physics reach of these configurations.

%-----------------------------------------------------------------------%

\section{Dark Matter}
\label{sec:dm}
The Lagrangian of interest for the analysis in this section is
\begin{equation}
\label{EqIII.1}
\mathcal{L} \supset \frac{1}{2}m_{A^{'}}^2A^{'}_\mu A^{'\mu} - m_\chi^2|\chi|^2 + ig_DA^{'}_\mu J_\chi^{\mu} + \epsilon e Z_{\mu}^{'}J^{\mu}_{EM} 
\end{equation}
where, $J_\chi^{\mu} = i(\chi^{*}(\partial^\mu \chi - (\partial^\mu \chi^{*})\chi)$ and $J^\mu_{EM} = \sum_{f} Q_f\bar{f}\gamma^\mu f$. We investigate dark matter of mass $m_\chi$ that couples to dark photons of mass $m_{A'}$ with a coupling strength $g_D$. We take scalar dark matter as a representative example in our analysis. The results can be straightforwardly extended to fermionic dark matter with similar sensitivity. We express the sensitivity in terms of $Y = \epsilon^2 \alpha_D (m_{\chi}/m_{A'})^4$ where $\epsilon$ is the coupling between $A'$ and charged SM fermions, and $\alpha_D = \frac{g_D^2}{4\pi}$. Fig.~\ref{fig:beam_dump} shows the sensitivity at the $90\%$ confidence level for a vector mediator model with $m_{A^{'}}=3m_\chi$ and a dark-sector coupling $\alpha_D=0.5$. The five beam configurations: $\nu$-mode, off-target mode with 1-, 2-, and 3-year exposures, and a dedicated 3-year beam-dump run correspond to the operational scenarios introduced at the beginning of Sec.~\ref{sec:beamdump} and listed in Tab.~\ref{tab:ExperimentalDetails_1}. In all beam configurations, the on-shell vector mediator ($A^{'}$) is produced via decays of neutral mesons or proton bremsstrahlung at the respective locations, promptly decaying into a pair of $\chi\bar{\chi}$. These DM particles propagate to the SBND detector and scatter off electrons or nucleons, producing signals through the DM-electron elastic scattering process $\chi e^- \rightarrow \chi e^-$, DM-proton elastic scattering process $\chi p \rightarrow \chi p$, and inelastic $N \chi \rightarrow N \chi \pi^0$ process. Below $m_\chi = 0.045~$GeV, DM is dominantly produced by $\pi^0$ decays. For $m_\chi \geq 0.045~$GeV, proton bremsstrahlung becomes the dominant production mechanism, with a minor contribution from $\eta$ decays.

\par In terms of detection, up to $m_\chi = 0.05~$GeV, the DM-electron elastic scattering channel plays the dominant role. Beyond this mass value, the proton elastic channel gradually becomes more significant. Both detection channels cease at $m_\chi$=0.5 GeV, as the 8 GeV proton beam does not produce enough dark photons via proton bremsstrahlung when $m_{A^{'}} \geq 1.5~$GeV. For a given set of masses and couplings, the inelastic $\pi^0$ channel is suppressed by $\mathcal{O}(1)$ as compared to the corresponding elastic channel in the cross-section and hence the number of events~\cite{deNiverville:2016rqh}. 

\par In Fig.~\ref{fig:beam_dump}, we show the sensitivity to the scalar DM for both electron and proton scattering channels. The existing constraints include BaBar $\gamma+$ missing energy search \cite{BaBar:2008aby, BaBar:2017tiz}, limits from electron recoil search in LSND \cite{LSND:2001akn, deNiverville:2011it}, MiniBooNE \cite{MiniBooNE:2017nqe}, E137 \cite{Batell:2014mga, Bjorken:1988as}, COHERENT CsI nuclear recoil searches \cite{COHERENT:2021pvd} and missing energy events in NA64 \cite{Banerjee:2019pds}. The black line represents the parameters that satisfy the thermal relic abundance~\cite{komatsu2009five} for scalar DM.

SBND can also probe fermionic DM, with cross sections similar to those of scalar dark matter. However, since we need larger couplings for the scalar scenario to probe the relic abundance~\cite{komatsu2009five}, we focus on scalar dark matter in this study. Apart from DM-electron and proton recoil and DM-inelastic $\pi^0$ production, SBND can probe sub-GeV DM through various other processes.
\vspace{0.5cm}

{\it{Other Dark Matter detection channels:}}\begin{itemize}
\item Detection via $e^+e^-$, $\mu^+\mu^-$ final states via dark matter internal pair production (DIPP)~\cite{Dutta:2024nhg} $2\rightarrow 4$ scattering. These channels provide opportunities to investigate flavor-dependent dark sector portals such as vector mediators appearing in $U(1)_{B-3L_{\mu}}$~\cite{Heeck:2018nzc}, $U(1)_{T_{3R}}$~\cite{Dutta:2019fxn}, etc.  
\item Detection via $\gamma$, $e\gamma$ final states. These final states emerge from $2 \rightarrow 3$ DM scattering with nucleus/electrons, such as $\chi+N\rightarrow\chi+N+\gamma$, $\chi+e\rightarrow \chi+e+\gamma$, etc. This provides an opportunity to probe DM via scalar portals that interact only with photons. 
\item Light dark matter can also be detected via inelastic nuclear scattering off argon nuclei, leading to the excitation of nuclear states. These excited nuclei subsequently de-excite through the emission of MeV-scale photons, producing localized low-energy “blip” signatures in a liquid argon time projection chamber. Running SBND in beam-dump mode can significantly improve the sensitivity in this channel by strongly suppressing neutrino-induced backgrounds~\cite{Dutta:2026zpe}.
\end{itemize}

All these new light DM search strategies and long-lived mediator solutions of the MiniBooNE/MicroBooNE anomaly will benefit tremendously from the off-target runs due to the higher $\pi^0$ and $\eta$ flux and less backgrounds due to the neutrinos in the ongoing target mode runs. The proposed BD based run would allow us to probe the dark sector and dark matter-based solutions~\cite{Dutta:2021cip, Dutta:new} in complementary channels, e.g.,  via $\pi^0.\,\eta$ decays etc.~\cite{CCM:2023itc}.

%-----------------------------------------------------------------------%

\section{Axion-Like Particles}
\label{sec:alp}
New pseudoscalar particles with couplings to the SM may exist as the remnants of broken global symmetries, in particular, heavy axions that are connected to solutions of the strong CP problem~\cite{Dine:1981rt,Zhitnitsky:1980tq,Kim:1979if,Shifman:1979if,PhysRevD.93.115010} and other axion-like particles~\cite{Gaillard:2018xgk,PhysRevLett.124.221801,Kivel:2022emq,PhysRevD.106.015030,Kitano:2021fdl,Valenti:2022tsc,Graham:2015cka,Banerjee:2020kww,Co:2022aav}.
They may be probed through gauge portal couplings, for example, or through effective couplings to the SM fermions~\cite{DiLuzio:2020wdo,Cicoli:2012sz,Sannino:2026wgx}.
These couplings have been probed at fixed target accelerator experiments via the numerous hadronic and electromagnetic showers created in the beam-dump~\cite{Fortin:2021cog}, where ALPs can be produced via numerous scattering channels.
Here we consider three simple benchmarks which lead to electromagnetic ($\gamma\gamma$ or $e^+e^-$) final states in the detector,
motivated by the advantage of an off-target mode run that evades potential neutrino-induced backgrounds (albeit they may be reducible by similar methods to those in refs.~\cite{Coloma:2023oxx,Brdar:2025hqi}). Additionally, for ALPs very near the neutral pion or eta masses, neutrino-induced coherent neutral meson production is an irreducible background, making the beam-dump mode uniquely sensitive in these windows~\cite{Brdar:2025hqi}.

(i) The first benchmark we consider, also investigated in refs.~\cite{Aloni:2018vki,Kelly:2020dda}, is that of \textit{gluon dominance}, in which we allow for an effective operator coupling the ALP to gluons,
\begin{equation}
\mathcal{L}_{\rm ALP-glu} \supset \frac{\alpha_s}{8\pi f_a} a G\Tilde{G}
\end{equation}
where $a$ is the ALP field, $\alpha_s$ is the strong coupling constant, and $G \Tilde{G} \equiv \rm{Tr}\, \epsilon_{\mu\nu\alpha\beta} G^{\mu\nu} G^{\alpha\beta}$ is the trace of the gluon field strength tensor and its dual.
The low energy effective field theory at one loop yields an effective photon coupling,
\begin{equation}
\mathcal{L}_{\rm ALP-glu}^{\rm 1-loop} \supset -\frac{1}{8\pi f_a} c_\gamma a F \Tilde{F}
\end{equation}
with
\begin{align}
c_\gamma &= -1.92 + \frac{1}{3} \frac{m_a^2}{m_a^2 - m_\pi^2} + \frac{8}{9} \frac{m_a^2 - \frac{4}{9} m_\pi^2}{m_a^2 - m_\eta^2} + \frac{7}{9} \frac{m_a^2 - \frac{16}{9} m_\pi^2}{m_a^2 - m_{\eta^\prime}^2}
\end{align}
for $m_a \lesssim \Lambda_{\rm QCD}$. This effective coupling allows us to look for $a \to \gamma \gamma$ decays inside the detector, with hadronic decays becoming dominant for $m_a \gtrsim 3 m_\pi$~\cite{Aloni:2018vki}. The ALP production in this case is supported through the gluon coupling giving rise to meson mixings $\theta_{a\pi}$, $\theta_{a\eta}$, $\theta_{a\eta^\prime}$ that allow for ALP production from the produced pion and eta mesons in the BNB target or dump. The production rate as a function of the mixings and the relevant factors accounting for kinematic phase space can be found in ref.~\cite{Kelly:2020dda}. 

We model the ALP production through these mixings utilizing simulations of the $\pi^0$ and $\eta$ meson production in the BNB target and dump in GEANT4. We then account for the ALP propagation from their production point into the SBND fiducial volume, where their probability to decay within the length of the detector is
\begin{equation}
    P_{decay} = e^{-\ell \Gamma_a / v_a} \big( 1 - e^{-\Delta \ell \Gamma_a / v_a} \big) \, ,
\end{equation}
where $\Gamma_a$ is the total ALP decay width in the laboratory frame, $\ell$ is the BNB target-SBND or BNB dump-SBND distance, and $\Delta\ell$ is the length of the fiducial volume along the beam axis. For the decays we will consider the full spectrum of final states; we have $a \to \gamma \gamma$ dominating at $m_a \lesssim 3 m_\pi$, and for higher masses primarily $a \to 3\pi$, $a \to \pi \pi \gamma$, and $a \to \eta \pi \pi$ (charged and neutral combinations). 

\begin{figure}
    \centering
    \includegraphics[width=0.5\textwidth]{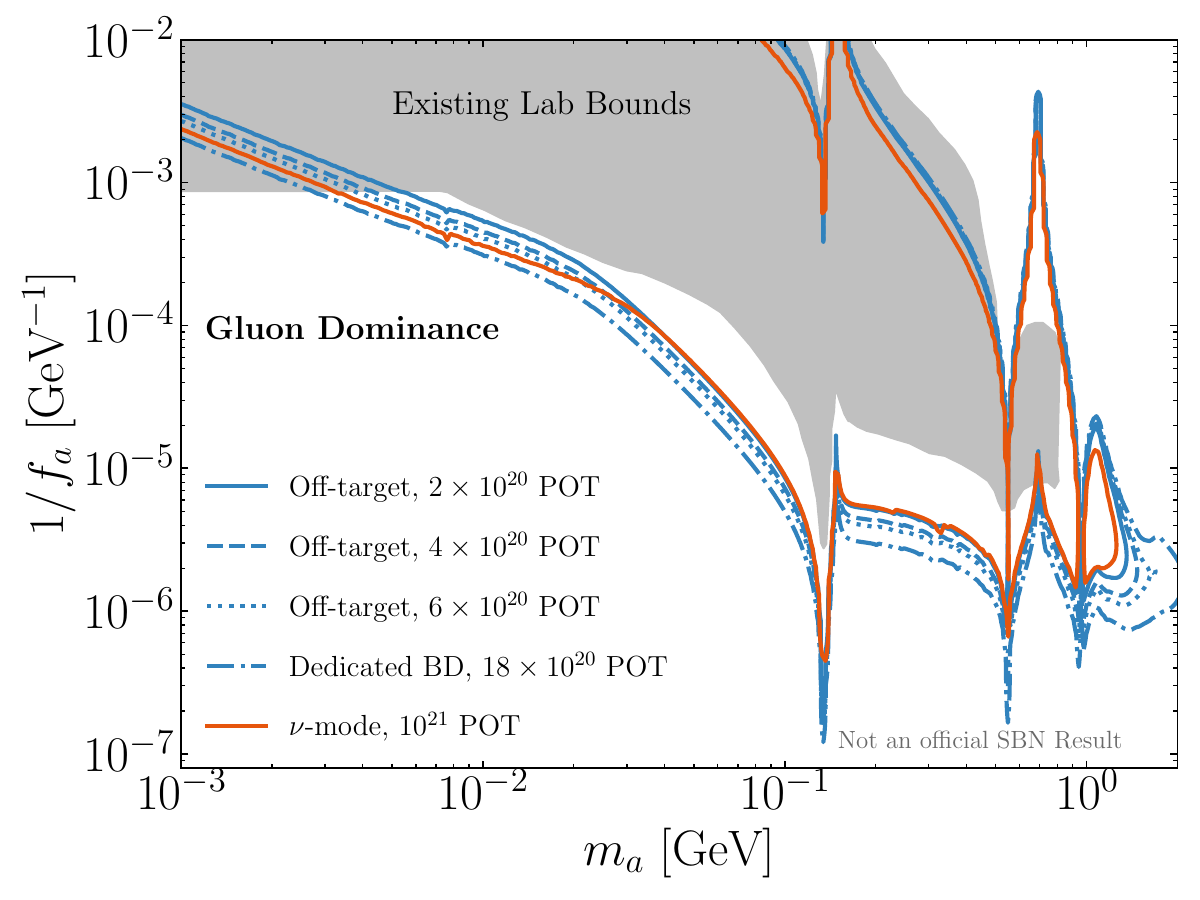}\\
    \includegraphics[width=0.5\textwidth]{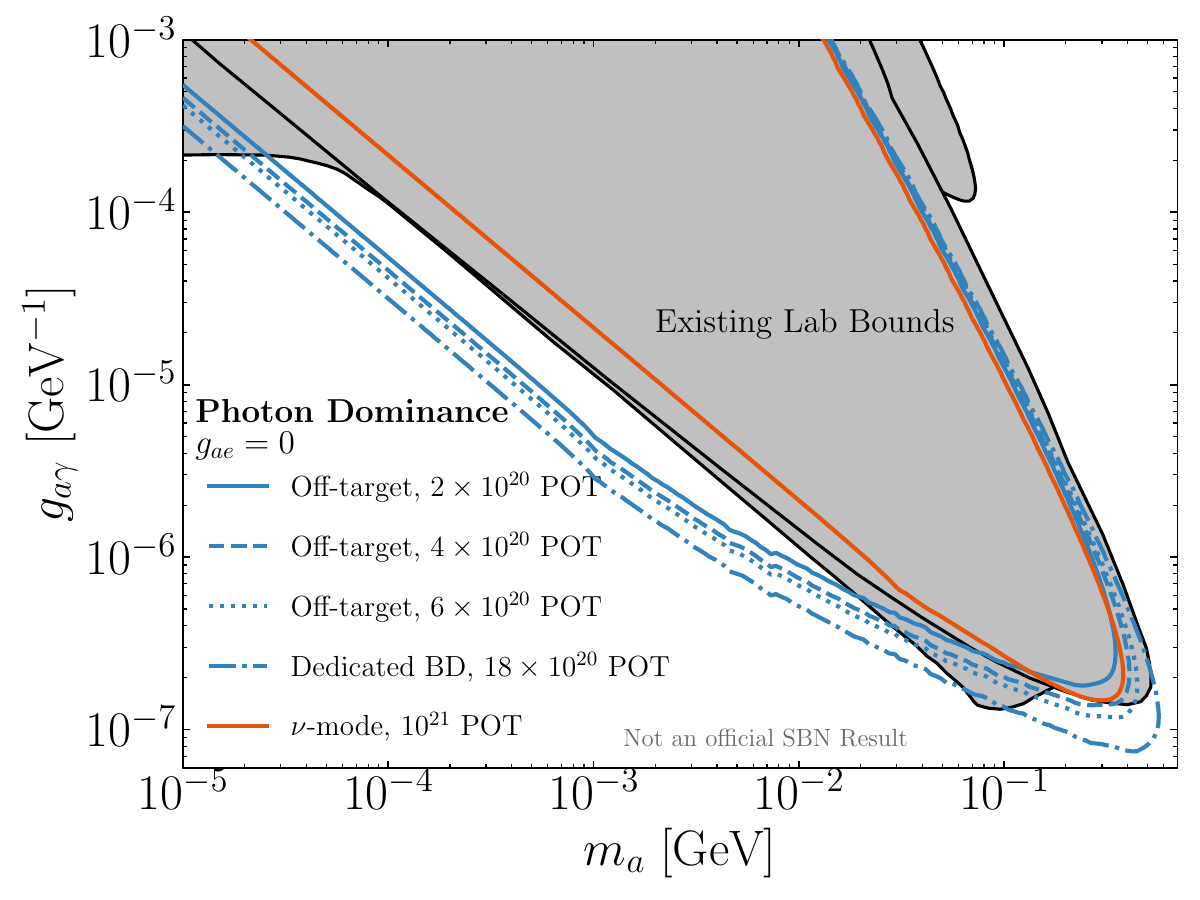}\\
    \includegraphics[width=0.5\textwidth]{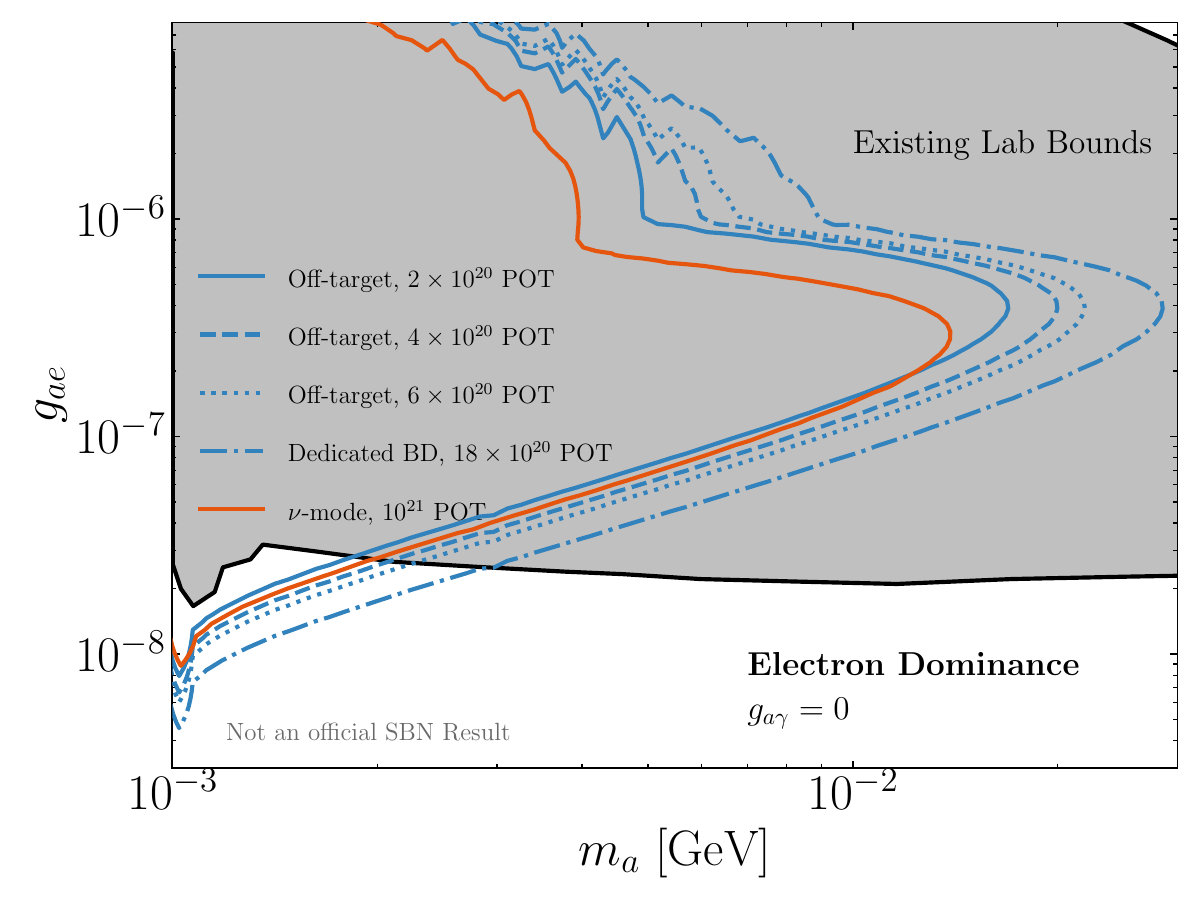}
    \caption{SBND sensitivity in Off-target beam-dump mode (blue contours) and in target neutrino mode running (orange contours) to axion-like particles under several benchmark model assumptions; \textit{top}: gluon dominance, \textit{middle}: photon-dominance, \textit{bottom}: electron-dominance.}
    \label{fig:alp_sens}
\end{figure}

With a final state reconstruction efficiency of $25\%$ assumed, the resulting sensitivity in the $(m_a, f_a^{-1})$ plane is shown in Fig.~\ref{fig:alp_sens} (top panel) for both target (neutrino) mode and off-target beam-dump running over the benchmark POTs we have assumed in this work. 
When the ALP mass is very close to the masses of the $\pi^0, \eta, \eta^\prime$, resonances in the mixings offer enhanced sensitivity to the parameter space. However, in the target mode the neutrino backgrounds from coherent neutral meson production is irreducible, making the dump mode advantageous in these mass windows.

(ii) Secondly, we consider a simplified scenario in which the ALP only couples to photons with negligible couplings to other particles. For this setup we assume the parameterization for the effective Lagrangian
\begin{equation}
    \mathcal{L}_{\rm{ALP}-\gamma} = -\frac{1}{4}g_{a\gamma} a F \Tilde{F} \, .
\end{equation}
In this phenomenological model, the signal is purely $2\gamma$ decays in the detector (with single photon scattering dominating only at masses $m_a \lesssim 10$ keV), and the ALP production takes place in the target through Primakoff scattering of electromagnetic cascade photons, thereby offering a phenomenology with softer fluxes and a simplified set of final states. The resulting sensitivity in the $(m_a, g_{a\gamma})$ plane is shown in Fig.~\ref{fig:alp_sens} (middle panel). These compare with constraints from existing beam-dump experiments (including E137, CHARM, nuCal, and BaBar~\cite{Dolan:2017osp, Aloni:2019ruo}, FASER~\cite{FASER:2024bbl}, and the MiniBooNE beam-dump mode~\cite{Capozzi:2023ffu}). The sensitivity to parameter space at smaller couplings offers a laboratory probe of bounds from astrophysics, namely by supernova~\cite{Caputo:2021rux,Fiorillo:2025sln} and stellar cooling~\cite{Hardy:2016kme}).

(iii) Lastly, we consider the scenario in which the ALP couples to electrons with negligible couplings to other particles, parameterized by the effective Lagrangian
\begin{equation}
    \mathcal{L}_{\rm{ALP}-e} = -i g_{ae} a \bar{e} \gamma^5 e \, .
\end{equation}
As investigated also in refs.~\cite{Patrone:2025fwk,Brdar:2025hqi}, this phenomenological coupling allows for production of ALPs in the beam-dump/target via electron/positron bremsstrahlung ($e^\pm Z \to e^\pm Z a$), resonant and associated production with a photon ($e^+ e^- \to a$, $e^+ e^- \to a \gamma$), and Compton scattering ($e^- \gamma \to a e^-$). ALPs are then free to propagate to the detector once produced, taking into account their lifetime. For $m_a > 2 m_e$, decay to electron-positron pairs is allowed and dominates the signal characterization.

The resulting sensitivity in the $(m_a, g_{ae})$ plane is shown in Fig.~\ref{fig:alp_sens} (bottom panel) which we compare with bounds from beam-dump bounds in gray~\cite{PhysRevD.38.3375,Andreas:2010ms,Bechis:1979kp,Riordan:1987aw,Bross:1989mp,NA64:2021ked,Andreev:2021fzd,Gninenko:2017yus,CCM:2021jmk}.

%-----------------------------------------------------------------------%

\section{Heavy Neutral Leptons}
\label{sec:hnl}

\begin{figure}
    \centering
    \includegraphics[width=0.99\linewidth]{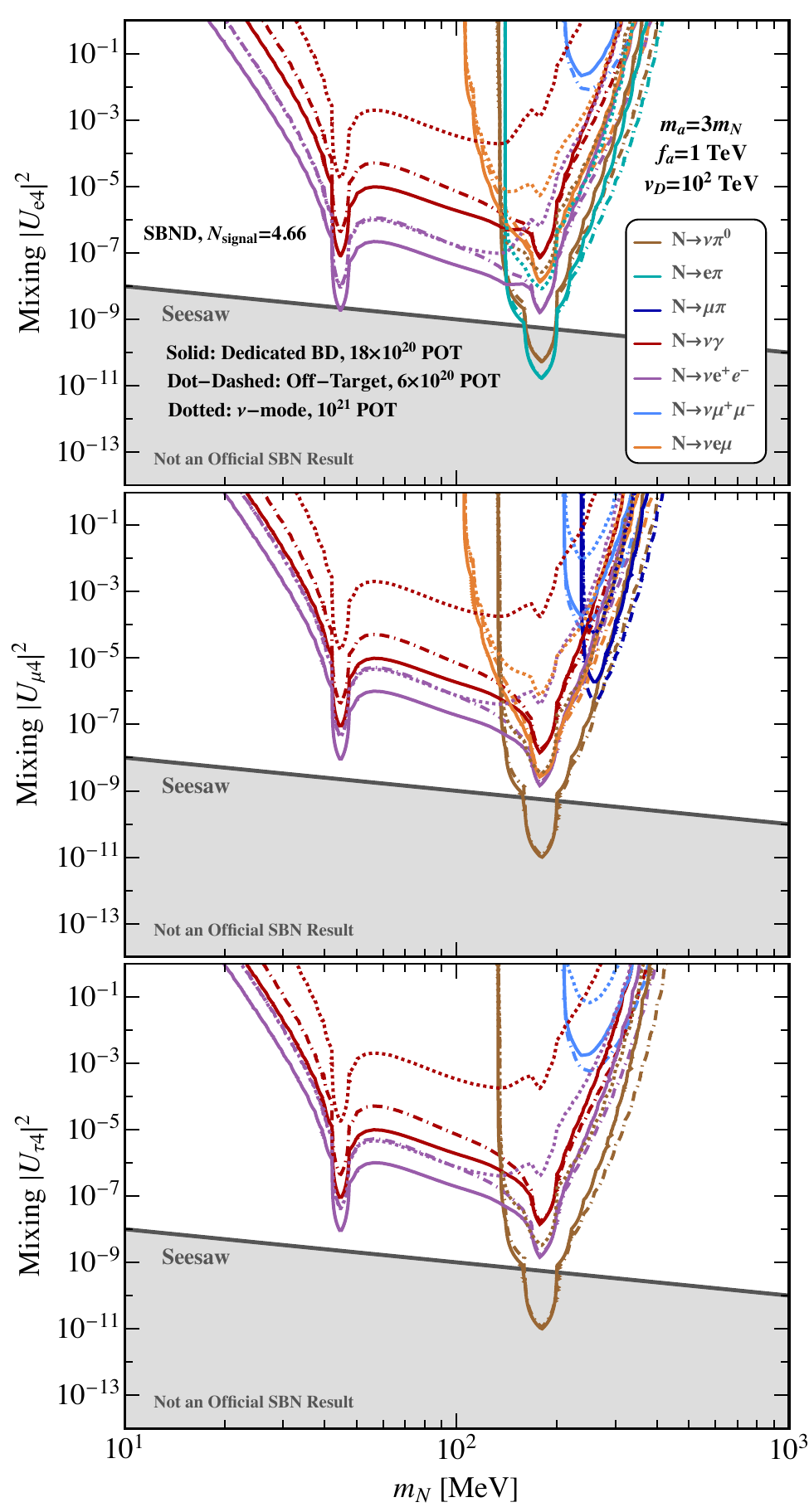}
    \caption{ALP-HNL sensitivity for all three scenarios, including background for the decay into $\nu \gamma$ for the neutrino mode.}
    \label{fig:HNLBcg}
\end{figure}

All different scenarios considered for SBND can explore Heavy Neutral Leptons (HNLs) as well, where $N$ interact with the SM only through mixing with active neutrinos $U_{\alpha4}$. Traditionally, it is assumed that HNLs are produced and detected via the same mechanism, which means the production of neutrinos would be proportional to $|U_{\alpha4}|^2$ times the fluxes of the SM neutrinos, while the detection would also be suppressed by $|U_{\alpha4}|^2$. This means that overall, the expected rate of HNLs at the detector would be suppressed by $|U_{\alpha4}|^4$. 

\begin{figure*}
    \centering
    \includegraphics[width=0.495\textwidth]{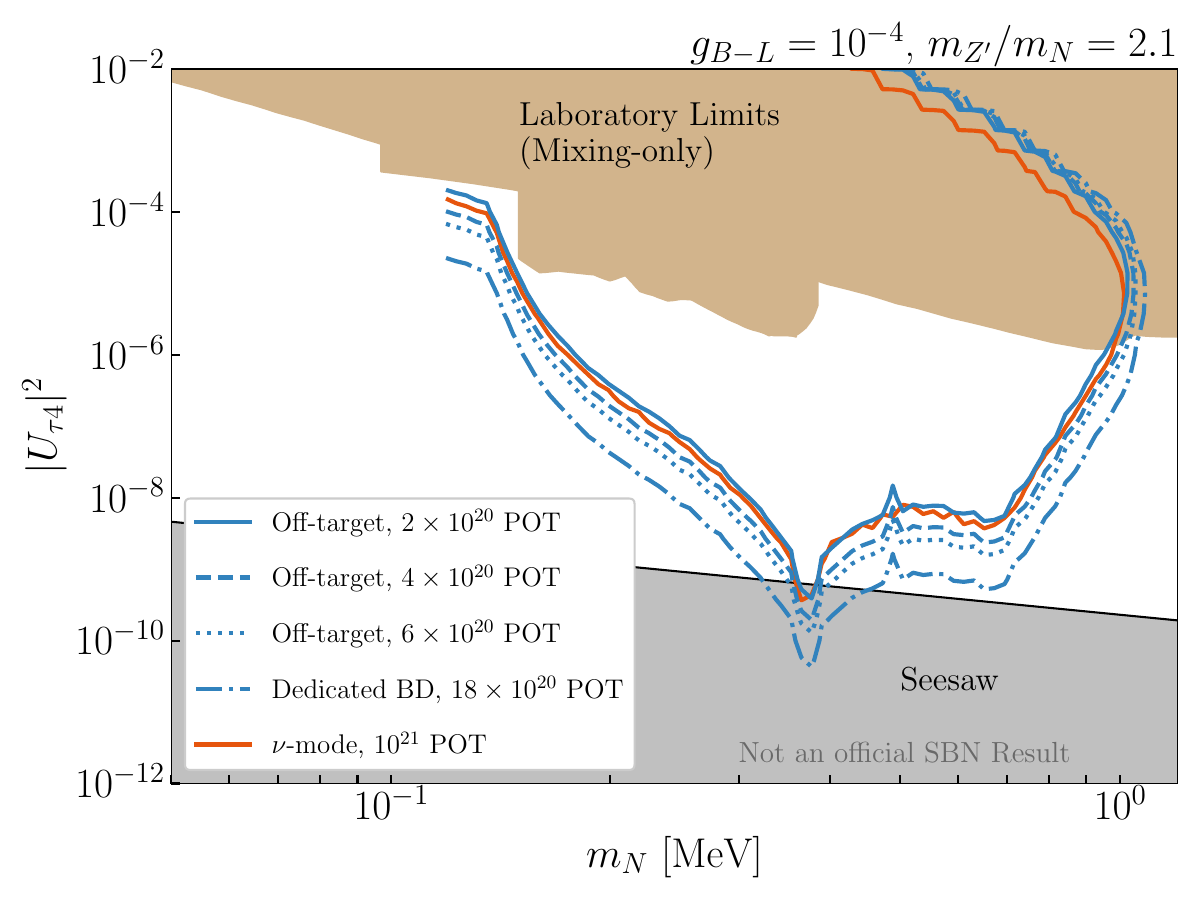}
    \includegraphics[width=0.495\textwidth]{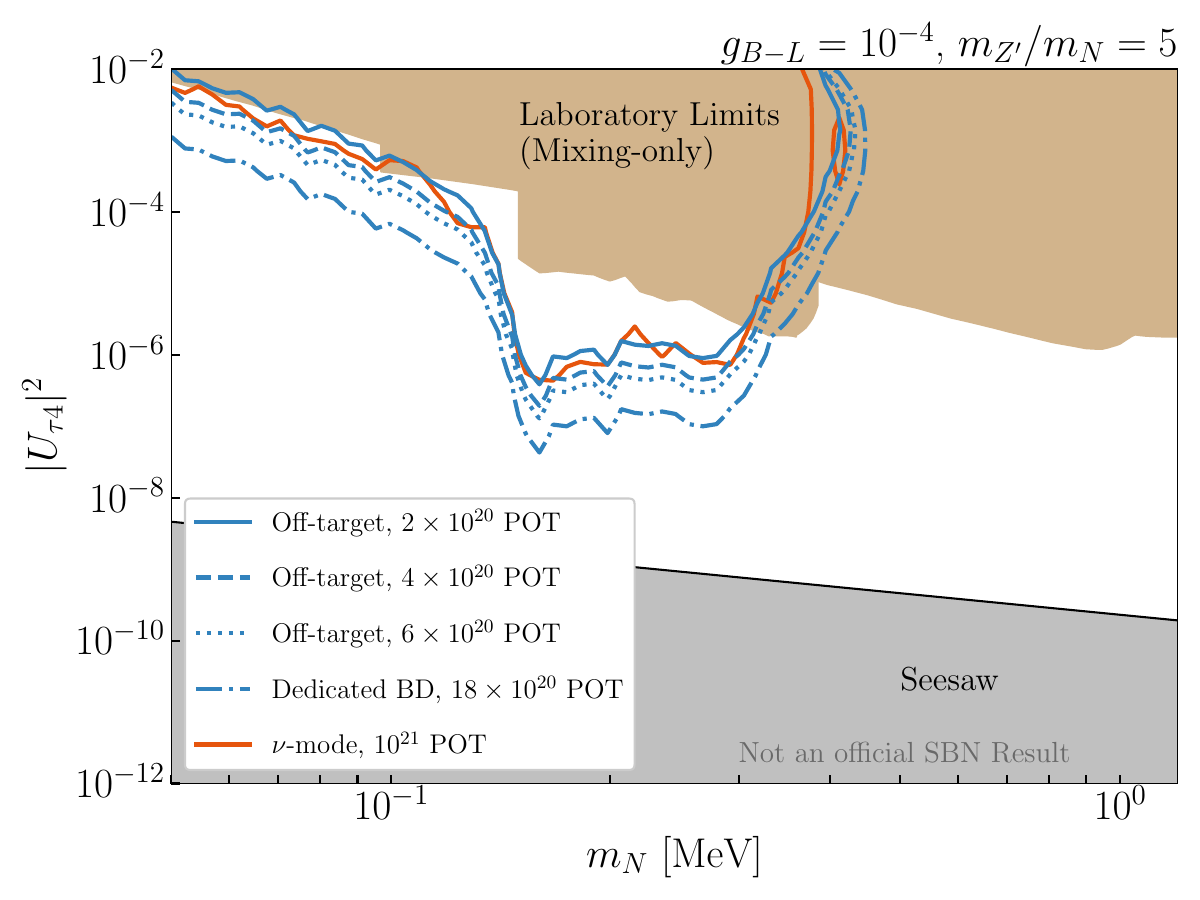}
    \caption{Sensitivity to the tau flavor HNL mixing $|U_{\tau4}|^2$ via the $B-L$ assisted HNL production, taking $m_{Z^\prime}/m_N = 2.1$ (\textit{left}) and $m_{Z^\prime}/m_N = 5$ (\textit{right}). The sensitivity curve on the right plot terminates due to a limitation imposed on $m_{Z^\prime}$ in the calculation of proton bremsstrahlung; see ref.~\cite{Capozzi:2024pmh} for details.}
    \label{fig:hnl_bl}
\end{figure*}

\textit{\textbf{Scenario I: HNLs interacting with an axion sector}}: However, as the authors of Ref.~\cite{Abdullahi:2023gdj} have shown, the new SM singlet fermions can belong to a much richer dark sector, which has its own particle contents, local symmetries and interactions. In this case, the HNLs, while they are still singlet under the SM, can be charged under the dark sector and interact with the dark sector particles. In that work, it is assumed that the dark sector consists of new pseudo-scalar axion-like particles (ALPs) $a$, and HNLs are predominantly produced via decays of ALPs, leading to potentially observable signals even for HNLs with minimal mixing. In this case, ALPs can be produced by mixing with the neutral SM mesons $\pi^0$ and $\eta$. The ALP would then travel towards the detector, and it can decay into a pair of HNLs. The HNLs that reach the detector may decay into the SM particles and yield an observable signal. In this case, the expected rate at the detector would be proportional to $\theta_{\mathbf{m}a}^2 |U_{\alpha 4}|^2$, where $\theta_{\mathbf{m}a}$ is the mixing between the SM meson and ALP particle, that depends on their masses and decay constants, and in principle we can have $\theta_{\mathbf{m}a}\gg U_{\alpha 4}$, and hence regions of the HNL parameter space that were previously inaccessible in fixed target experiments can be explored, including regions that are preferred by the type-I seesaw mechanism. A UV complete example of such dark sector can be found e.g. in Ref.~\cite{Berryman:2017twh}. However, we would like to emphasize that the results found here are general and they do not depend on any specific model. 

We perform an analysis of HNL produced via ALPs using SBND. In Fig~\ref{fig:HNLBcg}, we present the sensitivity of SBND to the HNL parameter space for each of the these assumed scenarios, where we have fixed some of the parameters of the dark sector model. As the three panels show, regardless of which neutrino flavor the HNL couples to, the overall sensitivities are approximately the same. This is because of separating the production and detection mechanism of ALPs, which is in contrast to the standard HNL scenarios where the production is via the mixing with the active neutrinos, and hence is proportional to the relative sizes of the fluxes of the charged meson parent particles. And as in these experiments we usually have $\phi_{\nu_\mu}\gg \phi_{\nu_e}\gg \phi_{\nu_\tau}$, we expect that the sensitivity to HNL coupled to each flavor be comparable with the relevant fluxes. As a result, we always get a much weaker sensitivity for HNL coupled to $\nu_\tau$. However, in the case considered here, the production of HNL is via ALPs mixing with the SM neutral mesons, the production rate is not suppressed by the active-sterile mixing parameter, and we see that the sensitivities are similar for different flavors. 

\textit{\textbf{Scenario II: HNLs interacting with new gauge forces}}: A second scenario in which HNL production could be enhanced and probed uniquely with the beam-dump mode is when the presence of a new gauge force coupled to the neutrino sector supports new production channels. One such example is the gauging of the difference of baryon and lepton number, $U(1)_{B-L}$, whose spontaneous breaking could be responsible for the neutrino mass generation. Studied in ref.~\cite{Capozzi:2024pmh} (see also refs.~\cite{Berryman:2017twh,Dev:2023zts}), this scenario is heuristically captured by invoking a massive mediator $Z^\prime$ that couples to $B-L$, therefore coupling to the HNLs $\mathcal{L}_{int} \supset -g_{B-L} Z^\prime_\mu \bar{N} \gamma^\mu N$ and the SM fermions with a coupling $g_{B-L} Q_{B-L}$. The primary driver for enhanced HNL production in this case comes from bremsstrahlung of $Z^\prime$ off the primary proton beam scattering off target nuclei $A$, $p + A \rightarrow X + Z^\prime$, with the subsequent prompt decay $Z^\prime \rightarrow N N$ sourcing a flavor-blind HNL flux.

\begin{figure*}[ht!]
    \centering
    \includegraphics[width=.60\linewidth]{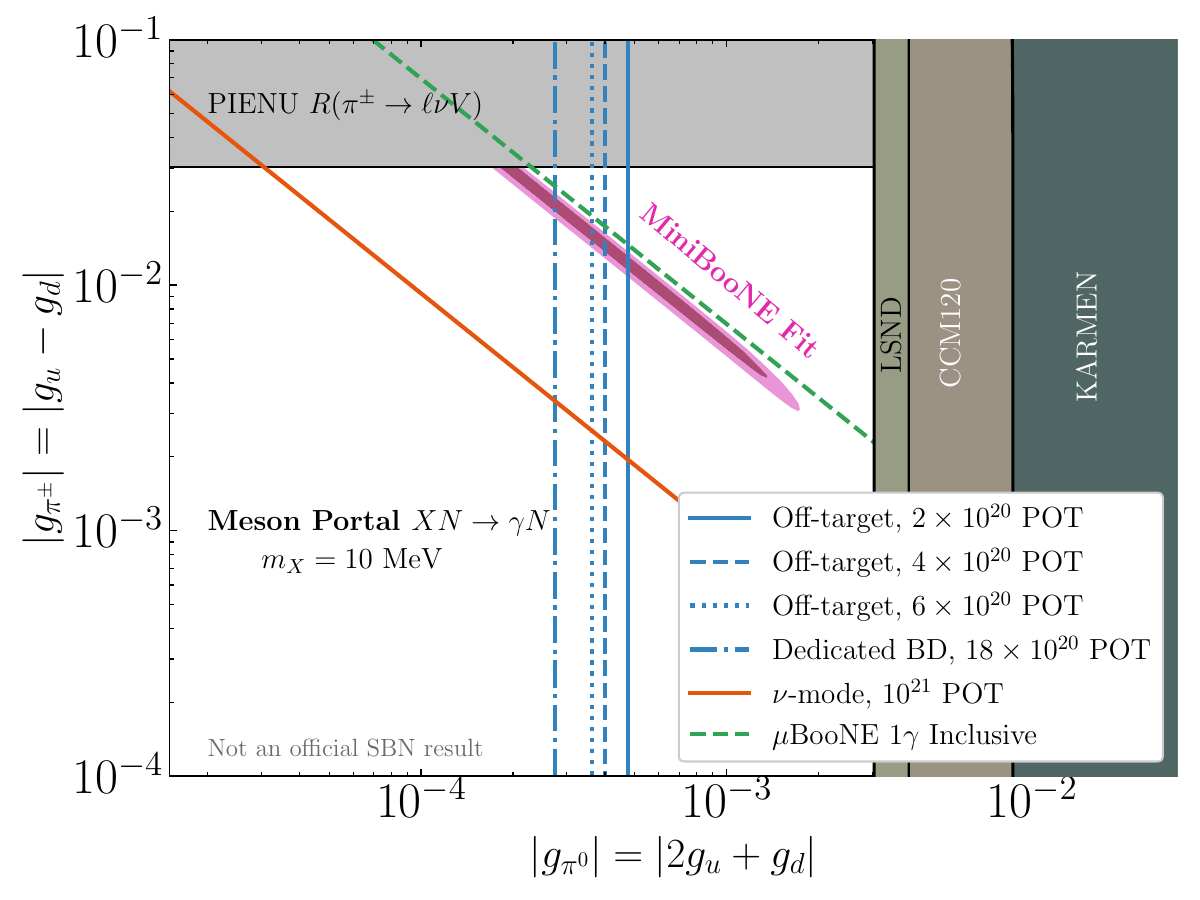}
    \caption{Parameter space limits for the meson portal vector $X$ for $m_X = 10$ MeV are shown over the vector neutral pion coupling ($g_{\pi^0}$) and charged pion coupling ($g_{\pi^\pm}$) are shown for the SBND off-target projections (blue lines) and the target mode running with $10^{21}$ POT (orange). The pink contours show the MiniBooNE best-fit regions ($1\sigma, 2\sigma$) as derived in ref.~\cite{CCM:2023itc}. Existing constraints are shown from stopped-pion experiments KARMEN, CCM, and LSND which are dominantly sensitive to the neutral pion decays $\pi^0 \to \gamma X$.}
    \label{fig:meson_portal}
\end{figure*}

The advantage of the dump mode to search for HNLs in this scenario lies partly in the flavor-blind production mechanism, allowing us to probe HNL decays $N \to \rm SM$ particles in a manner that is sensitive to all mixings, $|U_{e4}|^2$, $|U_{\mu4}|^2$, and $|U_{\tau4}|^2$. The second advantage is that the scattering of $Z^\prime$ in the proton bremsstrahlung process is characteristically forward, enhancing the flux acceptance of HNLs relative to the isotropic sources of active neutrinos that might ordinarily background processes. To evaluate the sensitivity to this scenario in the dump mode and make a comparison with the target mode, we adopt a similar set up as in ref.~\cite{Capozzi:2024pmh} and compute the rate of proton bremsstrahlung (using ref.~\cite{Foroughi-Abari:2021zbm}, see also refs.~\cite{Foroughi-Abari:2024xlj,Gorbunov:2024vrc, Gorbunov:2023jnx} for further discussion), inducing $Z^\prime$ production, with prompt decay to $NN$ pairs. We then compute the probability of HNLs $N$ propagating to the detector, taking into account their lifetime (driven primarily by the mixing $U_{\alpha 4}$) and flux acceptance. The less-constrained mixing to tau flavor can be probed in this way, setting $U_{e4} = U_{\mu4} = 0$ and $U_{\tau4} \neq 0$; the dominant final state in this case is the HNL decay $N \to \nu_\tau \pi^0$ (and $N\to \nu_\tau e^+ e^-$ below the pion mass threshold), ordinarily beset by neutrino backgrounds in the target mode but relatively background-free in the dump mode.

To explore the parameter space in a simplifying way, we fix the ratio of the $Z^\prime$ mass to the HNL mass, and fix $g_{B-L} = 10^{-4}$, showing the resulting sensitivity to $|U_{\tau4}|^2$ in Fig.~\ref{fig:hnl_bl} for the off-target and on-target benchmarks. We also show the existing bounds over $|U_{\tau 4}|^2$ that only considered HNL production due to the mixing for comparison; if these bounds were to be revisited, taking the presence of the $B-L$ gauge interactions into account, most of them would shift negligibly as argued in ref.~\cite{Capozzi:2024pmh}. Here, we find that when the $Z^\prime$ mass is sufficiently light (although limited to $m_{Z^\prime} > 2 m_N$ to keep $Z^\prime \to NN$ on-shell), the resulting decay probability for the HNL flux can peak at larger masses and smaller mixings, allowing the sensitivity across these benchmarks to approach and test the seesaw parameter space. The comparison with the $\nu$-mode and off-target modes is again put on equal footing here, assuming full rejection of neutrino-induced backgrounds at the cost of minimal signal loss in the $\nu$-mode, but even here the forecast favors better sensitivity from off-target running where the shorter baseline improves flux acceptance in the forward direction. This sensitivity, driven by the production of $Z^\prime$ from the proton beam, should be taken in complement with the HNL flux driven by lepton mixing and charged meson decays, but the parameter space shown here is only meant to capture the enhancement alone.

%-----------------------------------------------------------------------%

\section{Meson Portals}
\label{sec:meson}

Our searches for new physics in fixed target experiments are also well motivated by solutions to the short baseline anomalies, and in particular by solutions to the MiniBooNE excess. In light of the recent hints from MicroBooNE~\cite{MicroBooNE:2025ntu} mildly pointing to mono-photon signatures, for which there are both neutrino-based solutions~\cite{Harvey:2007ca,Harvey:2007rd,Gninenko:2010pr,Dutta:2025fgz,Kamp:2023hai} and solutions involving new propagating particles produced in the beam~\cite{Chang:2021myh, Fischer:2019fbw, Dutta:2021cip,CCM:2023itc,Dutta:2025fgz}. See also ref.~\cite{Abdallah:2022grs} and ref.~\cite{Kamp:2023hai} for a general review. Dedicated off-target running without an enhanced neutrino flux from meson focusing will help distinguish if this excess truly comes from unexplained physics in neutrino scattering cross sections, or if the anomaly is more correlated with neutral, unfocused particles in the beam target, as in the dedicated MiniBooNE beam-dump run~\cite{MiniBooNE:2017nqe}.

We consider here one phenomenological model involving a baryophillic mediator coupling to charged and neutral pions~\cite{Dutta:2021cip,CCM:2023itc}. These couplings may arise from direct couplings to up and down quarks that manifest as effective couplings to the charged pion current through the chiral perturbation theory, for example, parameterized with the effective Lagrangian
\begin{align}
\label{eq:singlemedLag}
   \mathcal{L}_{\rm X \pi} \supset & \, ig_{\pi^\pm} X_\mu \pi^+ (\partial^\mu \pi^-) + g_{\pi^0} \frac{e}{16\pi f_\pi}\pi^0 F_{\mu\nu}\Tilde{H}^{\mu\nu} \nonumber \\
   &- ig_{\pi NN} \pi^0 \overline{N} \gamma^5 \tau_3 N \, .
\end{align}
The dimensionless charged pion coupling, $g_{\pi^\pm}$, and the anomalous neutral pion coupling, $g_{\pi^0}$, involving the vector $X$ and field strength $\Tilde{H}^{\mu\nu} = \epsilon^{\mu\nu\alpha\beta}\partial_{[\alpha}X_{\beta]}$ and electromagnetic field strength $F_{\mu\nu}$, supports production of $X$ in the beam-dump via
\begin{align}
    \pi^0 & \to \gamma X \, , \nonumber \\
    \pi^\pm & \to \ell^\pm \nu_\ell X \, , \nonumber
\end{align}
and subsequent detection by neutral pion exchange with detector nucleons through the mono-photon production channel,
\begin{align}
    X N &\to \gamma N  \, .\nonumber
\end{align}

The resulting sensitivity over the $(g_{\pi^0}, g_{\pi^\pm})$ parameter space in the off-target mode running and target mode running for SBND is shown in Fig.~\ref{fig:meson_portal}. Existing bounds obtained from stopped pion experiments LSND~\cite{LSND:2001akn,LSND:2001fbw,LSND:1997vqj}, KARMEN~\cite{Karmen199815}, and CCM120~\cite{CCM:2023itc} are dominated by the $\pi^0 \to \gamma X$ channel and therefore limit the neutral pion coupling $g_{\pi^0}$. Explicit searches for missing energy in charged pion decays $\pi^\pm \to \ell \nu_\ell X$ for $\ell = e, \mu$ at PIENU, on the other hand, independently constrain the charged pion coupling $g_{\pi^\pm}$~\cite{PIENU:2021clt}.

%-----------------------------------------------------------------------%

\section{Summary}
\label{sec:summary}

In this work, we have examined the new physics opportunities enabled by operating the booster neutrino beam at Fermilab in an off-target and a dedicated beam-dump configuration to enhance the sensitivity of the SBND experiment to dark sector physics. By redirecting the proton beam away from the nominal Be target, neutrino-induced backgrounds can be suppressed by approximately a factor of 50 relative to standard neutrino-mode operation. In a dedicated beam-dump mode configuration, background suppression can reach up to three orders of magnitude. These reductions creates a significantly cleaner environment for searches involving final states such as $e$, $\gamma$, $\gamma\gamma$, and $\pi^0$, which are otherwise subject to substantial neutrino-induced backgrounds.

We have demonstrated that off-target running substantially extends the projected sensitivity of SBND to a broad class of dark sector scenarios. As representative examples, we presented sensitivities to dark matter, axion-like particles, heavy neutral leptons, and meson-portal models motivated in part by short-baseline anomalies. These studies highlight the unique capability of combining a high-intensity proton beam with a near, high-resolution liquid argon time projection chamber to probe light mediators and weakly interacting states.

Off-target operation also provides a realistic and technically accessible first step toward a more dedicated beam-dump program. The experience gained in such running modes, together with continued accelerator upgrades under PIP-II, would pave the way for a purpose-built beam-dump configuration capable of achieving even stronger neutrino suppression and enhanced neutral meson production. Such a program would significantly broaden the physics reach of SBND.

More broadly, these potential operation scenarios at SBND offers a cost-effective and complementary probe of light dark sectors, operating in synergy with underground direct-detection experiments, astrophysical observations, neutrino telescopes, and collider searches. In this context, off-target and dedicated beam-dump configurations at SBND provide a compelling opportunity to explore previously inaccessible regions of parameter space in the next few years.

\section*{Acknowledgements}
We thank Wooyoung Jang and Hyunyong Kim for their work on \texttt{GEANT4} simulations. The work of BD, DG, and AK is supported in part by the U.S. Department of Energy Grant~DE-SC0010813.  The work of ZT is supported by Pitt PACC and CERN's Theoretical Physics department. The work of AT is supported in part by the U.S. DOE grant DE-SC0010143. This manuscript has been authored by FermiForward Discovery Group, LLC under Contract No. 89243024CSC000002 with the U.S. Department of Energy, Office of Science, Office of High Energy Physics. \\
Note: The work and conclusions presented in this publication are not to be considered as results from the SBN Collaboration.

\appendix

%-----------------------------------------------------------------------%

\section{Meson and Photon Fluxes}
\label{sec:fluxes}

\begin{figure}
    \centering
    \includegraphics[width=0.95\columnwidth]{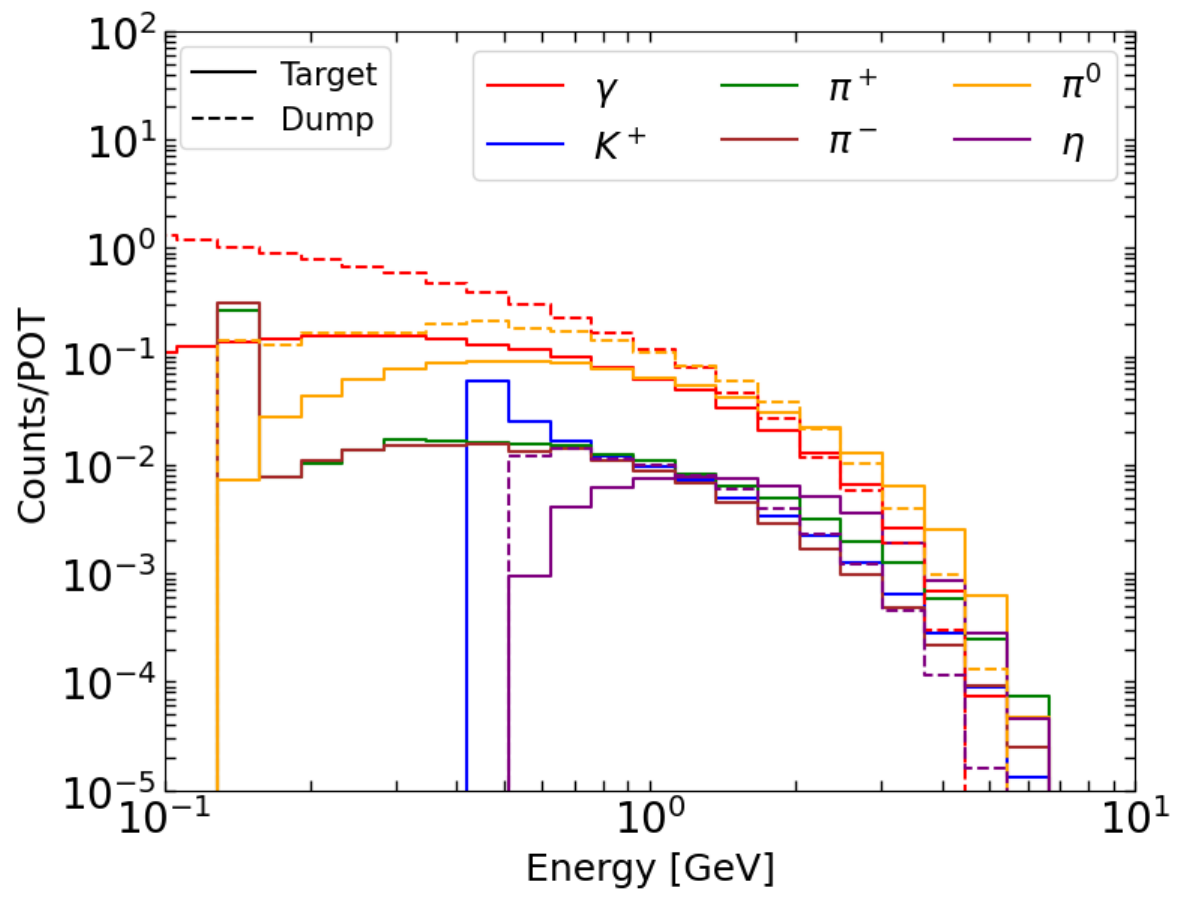}
    \caption{Particle yields per POT in the BNB $^9$Be target (solid) and the $^{56}$Fe dump (dashed), respectively. The fluxes are simulated~\cite{Dutta:2026zpe, Dutta:2025sba, Capozzi:2021nmp, Capozzi:2024pmh, Dutta:2025npn} using the \texttt{GEANT4} package.}
    \label{fig:fluxplot}
\end{figure}

We present a comparison of photon and meson ($\pi^\pm$, $\pi^0$, $\eta$, and $K^+$) production rates at the $^9$Be target and the $^{56}$Fe dump in Fig.~\ref{fig:fluxplot}. The quoted yields correspond to particle production at the primary interaction point per POT. Charged mesons will decay thereafter. In Tab.~\ref{tab:ExperimentalDetails}, we summarize different particle production per POT at the $^9$Be target and $^{56}$Fe dump. 

\begin{table*}[!htbp]
    \centering
    \renewcommand{\arraystretch}{2.0}
%     \begin{tabular}{ | c | c | c | c | c | }     
    \begin{tabular}{ | c | c | c | c | c | c | c |}
        \hline
         Target & $\gamma$ per POT  & $\pi^0$ per POT & $\eta$ per POT & $K^+$ per POT & $\pi^+$ per POT & $\pi^-$ per POT \\
        \hline
        $^9$Be & 3.645 & 0.883 & 0.052 & 0.143 & 0.431 & 0.456 \\ 
        \hline
        $^{56}$Fe & 2227.759 & 1.997 & 0.070 & - & - & - \\ 
        \hline
    \end{tabular}
    \caption{Comparison of different particle production per POT between the $^9$Be target and $^{56}$Fe dump of BNB~\cite{Dutta:2026zpe, Dutta:2025sba}.} 
    \label{tab:ExperimentalDetails}
\end{table*}

\section{Background Calculations}
\label{sec:bkgcal}
In Ref.~\cite{MicroBooNE:2015bmn}, the number of proton and $\pi^0$ background events is reported as 1,371,070 and 358,443, respectively, for $6.6\times10^{20}$ POT in target mode, which corresponds to our $\nu$-mode configuration (with different POT). To obtain the correct background counts for our $\nu$-mode, these values are scaled by the ratio of POT between the two modes, i.e., by a factor of 1.52. The same scaling applies to both the proton and $\pi^0$ backgrounds.  

For other operational scenarios, the background counts are estimated using the following relation:
\begin{equation}
    N_{\text{bkg}} = N_{\nu\text{Mode}} \times \frac{\text{POT}}{10^{21}} \times \left(\frac{110}{l_d}\right)^2 \times \frac{1}{\text{Factor}},
\end{equation}
where $N_{\text{bkg}}$, POT, and $l_d$ denote the background count, protons on target, and detector distance from the target or dump, respectively, for the scenario considered. The ``Factor" takes values of 50 and 1000 for the ``Off-Target Mode" and ``Dedicated Beam-Dump Mode," respectively, to scale the backgrounds appropriately.

\section{Recoil Energy Plots}
\begin{figure*}
    \centering
    \includegraphics[width=0.48\linewidth]{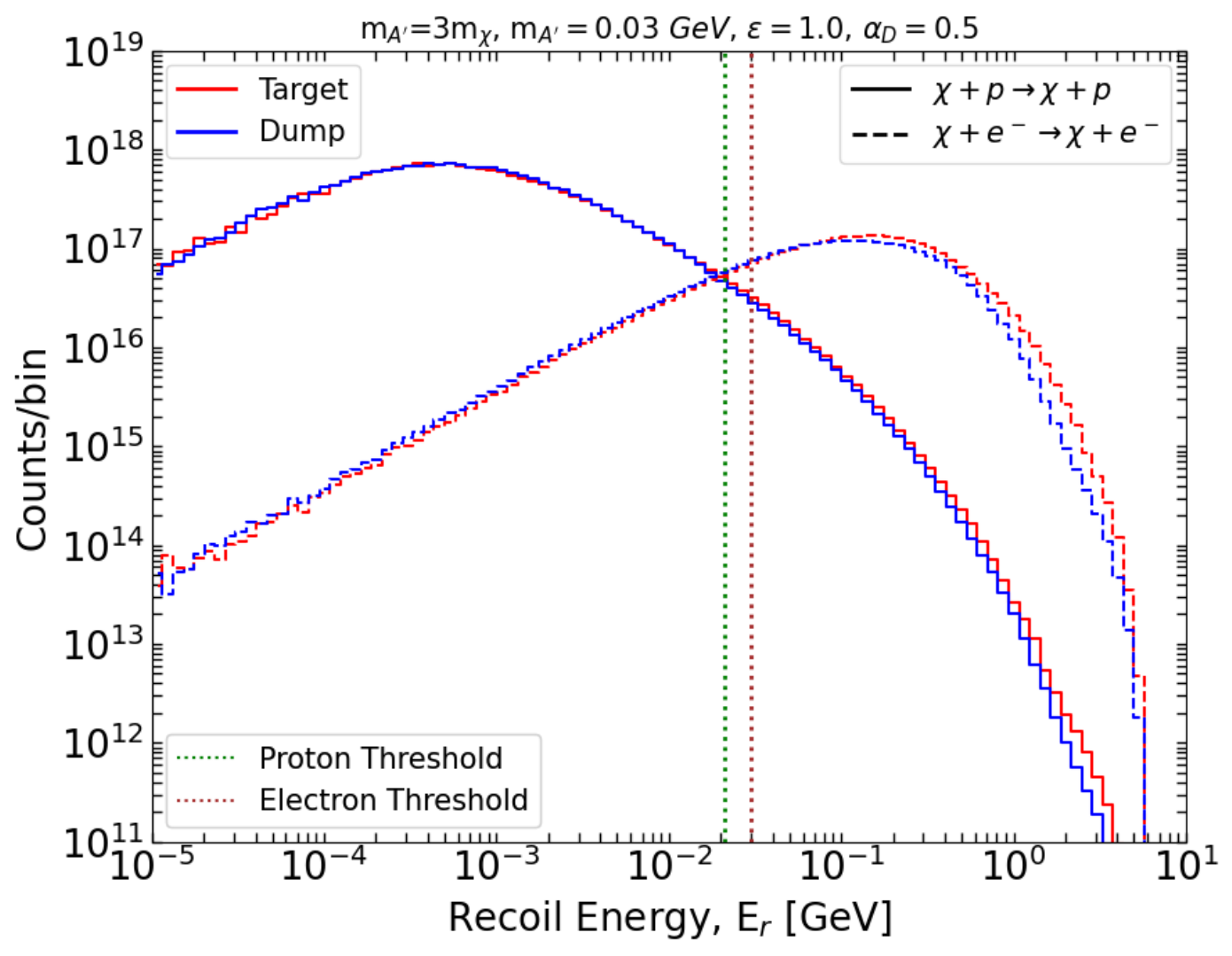}
    \includegraphics[width=0.48\linewidth]{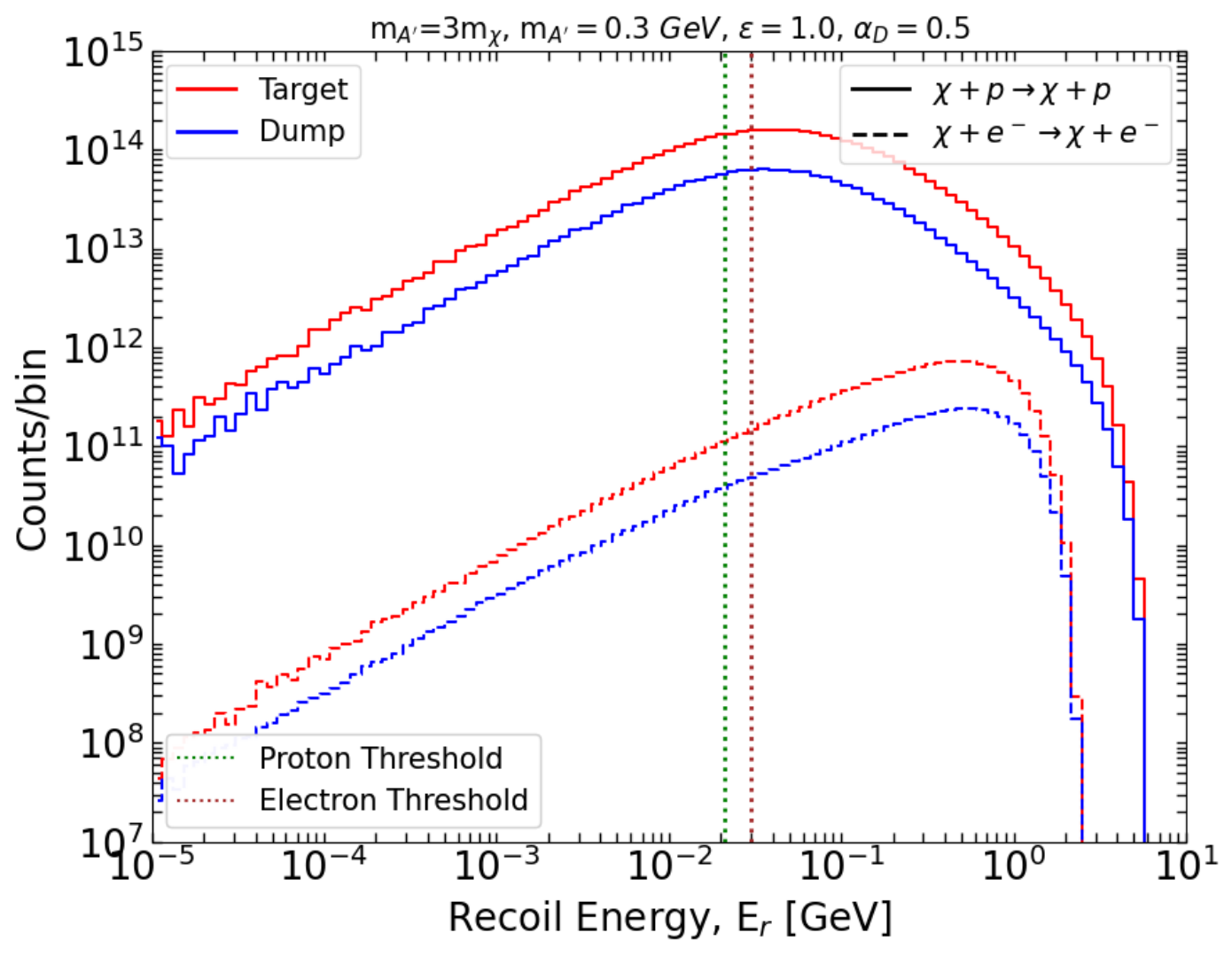}
    \caption{Recoil energy for m$_{A^{'}}=0.03~$GeV (left) and m$_{A^{'}}=0.3~$GeV (right).}
    \label{fig:protonelectron_recoil}
\end{figure*}

In Fig.~\ref{fig:protonelectron_recoil} we compare the recoil energy of proton and electron elastic scattering channels for selected benchmark points of $m_{A^{'}}=0.03~$GeV (left) and $m_{A^{'}}=0.3~$GeV (right) in the $\nu$-mode, the solid and dashed lines denote the target and dump contributions respectively. The values of $\epsilon$ and $g_{D}$ couplings have been set to 1.0 and 2.5, respectively. The green and brown dotted lines denote the SBND detector threshold of proton and electron recoil energy at 21~MeV and 30~MeV, respectively. Since electrons are much lighter than protons, we observe peaks at higher energies for the electron elastic scattering channel for both masses. Additionally, the t-channel process prefers higher momentum when the mediator mass is larger, and therefore, we observe the peaks being shifted towards higher energy as the mass of the mediator increases for both channels.  

\bibliography{references}

\end{document}